\documentclass{amsart}  
\usepackage{amssymb}  
\usepackage{graphicx}  
\usepackage{pictex}  
  
\newcommand{\RR}{{\mathbb R}}

\newtheorem{theorem}{Theorem}  
\newtheorem{lemma}{Lemma}[section]  
\newtheorem{prop}[lemma]{Proposition}  
  
\newtheorem{definition}[lemma]{Definition}  
\newtheorem{remarks}[lemma]{Remarks}  
\newtheorem{remark}[lemma]{Remark}  
\newtheorem{thm}[lemma]{Theorem}  
\newcommand{\tr}{{\mathrm{tr}}}  
\newcounter{smalllist}

\newcommand{\CalP}{\mathcal{P}}

\newcommand{\Oomega}{(\Omega,T)}

\newcommand{\diam}{\mbox{diam}}

\begin{document}  
\title[Discontinuities of the integrated density of states]{  
 Discontinuities of the integrated density of states  
 for random operators on Delone sets  
  }  
  
\author[S. Klassert, D. Lenz and P. Stollmann]{Steffen Klassert~$^{1,*}$,  
Daniel Lenz~$^{2,*}$ and\  
Peter Stollmann~$^{3,*}$} \thanks{* Research partly supported  
by the DFG in the priority program Quasicrystals}  
  
\maketitle  
  
\vspace{0.3cm}  
\noindent  
$^{1}$ Fakult\"{a}t f\"{u}r  
Mathematik, Technische  
Universit\"{a}t Chemnitz,  
D-09107 Chemnitz, Germany;  
E-mail:  S.Klassert@mathematik.tu-chemnitz.de,\\[0.1cm]  
  
\noindent  
$^{2}$ Fakult\"{a}t f\"{u}r  
Mathematik, Technische  
Universit\"{a}t Chemnitz,  
D-09107 Chemnitz, Germany;  
E-mail:  D.Lenz@mathematik.tu-chemnitz.de,\\[0.1cm]  
  
\noindent  
$^{3}$  Fakult\"{a}t f\"{u}r Mathematik, Technische Universit\"{a}t  
Chemnitz,  
D-09107 Chemnitz, Germany; E-mail:  P.Stollmann@mathematik.tu-chemnitz.de

\begin{abstract}  
Despite all the analogies with "usual random" models, tight  
binding operators for quasicrystals exhibit a feature which  
clearly distinguishes them from the former: the integrated density  
of states may be discontinuous. This phenomenon is identified as a  
local effect, due to occurrence of eigenfunctions with bounded  
support.  
\end{abstract}

\maketitle

\section{Introduction}  
In the present article we study the occurrence of discontinuities  
in the integrated density of states (IDS). For the special case of  
a tight binding model associated with the Penrose tiling the  
occurrence of this effect has been known since quite some time as  
witnessed for example in \cite{ATF,FATK,KF,KS}. We present two  
results. The first aims at showing that the occurrence of jumps in  
the IDS cannot be excluded by global assumptions concerning, e.g.  
ergodic or combinatorial properties. To this end we present a  
theorem saying that starting from some model of aperiodic order  
(phrased in the language of Delone sets) one can construct a model  
that is "basically the same" and gives rise to a tight binding  
operator with a discontinuity in the IDS. Here "basically the  
same" is cast in the notion of "mutually locally derivable" for  
Delone dynamical systems. We discuss this notion analogous to  
the respective notion for tilings found in \cite{BSJ}. In the  
construction we use that the laplacians on certain graphs have  
finitely supported eigenfunctions. It now becomes clear that it is  
the more complex structure of graphs in higher dimension that  
makes such a phenomenon possible. On lattices (and, consequently,  
in one-dimensional systems) such finitely supported eigenfunctions  
cannot occur.  
  
Our second theorem says that this is the only possibility to create a jump of the IDS,  
at least when starting from a reasonable Delone dynamical system.  
It is a consequence of a rather strong ergodic theorem in \cite{LS3}.  
This theorem states that the IDS is in fact the uniform limit of eigenvalue counting  
distributions. It is therefore substantially stronger than the weak convergence results  
typically proven in connection with the IDS see for example \cite{BLT,H,K}. The fact that  
such a strong convergence holds true is special for models of aperiodic order  
and not met in usual random systems.

\section{Notation and results}  
In this section we introduce some notation and present our  
results. We will use the same setting as the one found \cite{LS2}.  
For completeness reasons we include the necessary definitions.  
  
\medskip  
  
Let $d\geq 1$ a fixed integer and  
all Delone sets, patterns etc. will be subsets of $\RR^d$. The  
Euclidean norm on $\RR^d$ will be denoted by $\|\cdot\|$. For $r\in  
\RR^+$ and $p\in \RR^d$, we let $B(p,r)$ be the closed ball in $\RR^d$ centered at  
$p$ with radius $r$.  
  
A subset $\omega$ of $\RR^d$ is called Delone set if there exist  
$r(\omega)$ and $R(\omega)>0$ such that $2r(\omega) \leq \|x-y\|$  
whenever $x,y\in \omega$ with $x\neq y$, and $B(x,R(\omega))\cap  
\omega \neq \emptyset$ for all $x\in \RR^d$.

We are dealing in this paper with local structures of Delone sets,  
therefore the restrictions of $\omega$ to bounded subsets of  
$\RR^d$ are of particular interest. In order to treat these  
restrictions, we introduce the following definition.

\begin{definition}{\rm (a)} A pair  $(\Lambda,Q)$ consisting of a bounded  
subset $Q$ of $\RR^d$ and  $\Lambda\subset  Q$ finite is called  {\rm pattern}.  
The set $Q$ is called the {\rm support of the pattern}. \\  
{\rm  (b)} A pattern $(\Lambda,Q)$ is called a {\rm ball pattern} if $Q=B(x,r)$  
  with $x\in \Lambda$ for suitable $x\in \RR^d$ and $r\in (0,\infty)$.  
\end{definition}  
  
The diameter and the volume of a pattern are defined to be the  
diameter and the volume of its support respectively.  
  
We will have to identify patterns which are equal up to  
translation. More precisely, on the set of patterns we introduce  
an equivalence relation by setting $(\Lambda_1,Q_1)\sim  
(\Lambda_2, Q_2)$ if and only if there exists a $t\in \RR^d$ with  
$\Lambda_1 = \Lambda_2 + t$ and $Q_1=Q_2 + t$. The class of a  
pattern $(\Lambda,Q)$ is then denoted by $[(\Lambda,Q)]$.  
Obviously the notions of diameter, volume occurrence etc. can  
easily be carried over from patterns to pattern classes.

Every Delone set $\omega$ gives rise to a set of pattern classes,  
$\CalP (\omega)=\{ Q\wedge \omega :  
Q\subset\RR^d\: \mbox{bounded and measurable} \}$, and to a set of  
ball pattern classes $\CalP_B (\omega) =\{[ B(p,r)\wedge \omega] : p\in  
\omega, r\in \RR^+ \}$.  Here we set $Q\wedge \omega= (\omega \cap Q,  
Q)$.  
We define the radius $s=s(P)$ of an arbitrary ball pattern $P$ to be the  
radius of the underlying ball.  
For $s\in (0,\infty)$, we denote by $\CalP_B^s (\omega)$ the set of  
ball patterns with radius $s$.  A Delone set is said to be of {\it finite  
type} or of {\it finite local complexity}  
if for every radius $s>0$ the set $\CalP_B^s (\omega)$ is finite.

The Hausdorff metric on the set of compact subsets of $\RR^d$  
 induces the so called {\it natural topology} on the  
set of closed subsets of $\RR^d$. It is described in detail  
in \cite{LS2} and shares some nice properties: firstly, the set of all  
closed subsets of $\RR^d$ is compact in the natural topology. Secondly, and  
this is of prime importance in view of the dynamical system we are to consider,  
the natural action $T$ of $\RR^d$  
on the closed sets in $\RR^d$ given by $T_t G= G+t$ is continuous.  
  
Furthermore, a {\it Delone dynamical system  (DDS)} consists of a  
set $\Omega$  
 of Delone  
sets, which is invariant under the shift $T$ and closed in the natural  
topology.  
A DDS is said to be {\rm of finite type (DDSF)} if  
$\cup_{\omega\in \Omega} P_B^s (\omega)$ is finite for every $s$ and  
the set $\CalP (\Omega)$ of patterns classes associated to a DDS  
$\Omega$ is defined by $\CalP (\Omega)=\cup_{\omega\in \Omega} \CalP  
(\omega)$.  
Due to the compactness of the set of all closed sets in the natural topology  
a DDS $\Omega$ is compact.  
We refrain from a precise discussion of the topology but we give the following  
Lemma from \cite{LS2}.  
\begin{lemma}\label{topology}  
If $\Oomega$ is a DDSF then a sequence $(\omega_k)$ converges to $\omega$  
in the natural topology if and only if there exists a sequence $(t_k)$  
converging to $0$ such that for every $L>0$ there is an $k_0\in\mathbb N$  
with $(\omega_k+t_k)\cap\ B(0,L)=\omega\cap B(0,L)$ for $k\geq k_0$.  
\end{lemma}  
Roughly speaking, $\omega$ is close to $\tilde\omega$ if $\omega$  
equals $\tilde\omega$ on a large ball up to a small translation.  
We now recall some standard notions from the theory of dynamical  
systems and some available equivalent ``combinatorial''  
characterizations.

A dynamical system $\Oomega$ is called {\it minimal} if the orbit  
$\{ T_t\omega : t\in\RR^d\}$ of any $\omega$ is dense in $\Omega$.  
For a DDS this is equivalent to the property that $\CalP (\Omega)=  
\CalP (\omega)$ for any $\omega$. This latter property is called  
{\it local isomorphism property} in the tiling framework; see  
\cite{Sol1}. A sequence $(Q_k)$ of subsets in $\RR^d$ is called a  
van Hove sequence if the sequence $|\partial^R Q_k| |Q_k|^{-1}$  
tends to zero for every $R\in (0,\infty)$. Here, $\partial^R Q$  
denotes the set of those $x\in \RR^d$ whose distance to the  
boundary of $Q$ is less than $R$. Furthermore, a dynamical system  
$\Oomega$ is called {\it uniquely ergodic} if it admits only one  
$T$-invariant measure (up to normalization). For a Delone  
dynamical system, this is equivalent to the fact, that for every  
nonempty pattern class $P$ the frequency  
$$ \nu(P)\equiv\lim_{k\to\infty}|Q_k|^{-1}\sharp_P(Q_k \wedge\omega), $$  
exists uniformly in $\omega\in\Omega$ for every van Hove sequence $(Q_k)$.  
We call a dynamical system $\Oomega$ {\it strictly ergodic} if it is  
minimal and uniquely ergodic.  
Note that in this case the frequency $\nu(P)$ is positive for every $P\in\CalP$  
  
\begin{definition} Let $\Oomega$ be a DDSF.  A family $(A_\omega)$ of bounded  
operators $A_\omega : \ell^2 (\omega) \longrightarrow \ell^2 (\omega)$  
is called a random operator of finite range if there exists a constant $r_A$ with  
\begin{itemize}  
\item $A_\omega (x,y) =0$ whenever $\|x-y\| \geq r_A$.  
\item $A_\omega (x,y)$ only depends on the pattern class $(B(x,r_A)\cup  
  B(y,r_A))\wedge \omega$.  
\end{itemize}  
  
\end{definition}  
  
\medskip  
  
Note that the above defined operators provide a framework  
including Laplace type operators defined on $\ell^2(\mathbb Z^d)$.  
Let us mention, that for $\tilde\omega,\omega\in\Omega$ the  
operators act on different Hilbert spaces $\ell^2(\omega)$ and  
$\ell^2(\tilde\omega)$ unless $\omega$ and $\tilde\omega$ differ  
only by translation. Thus, to deal with operators on  
$\ell^2(\omega)$ is more complicated than in the lattice case.  
  
\medskip  
  
The aim of this article is to discuss the phenomenon of  
discontinuities of the integrated density of states of random  
operators $(A_\omega)$ on a DDSF $\Oomega$. This might rather come  
as a surprise in view of what is known for random models as well  
one dimensional quasicrystals. It turns out that this phenomenon  
occurs if and only if there exist locally supported eigenfunctions  
of $(A_\omega)$. One can find examples of locally supported  
eigenfunctions on the Penrose tiling in \cite{KS} and \cite{ATF}.  
An eigenfunction $f$ is said to be locally supported if  
$\mathrm{supp} f\subset K$, with $K$ a compact set. The phenomenon  
of locally supported eigenfunctions is by no means pathological.  
Rather from any given DDSF $\Oomega$ we can construct an in some  
sense local equivalent DDSF $(\Omega^b,T)$ such that a random  
operator of finite range $(A^b_\omega)$ defined on $(\Omega^b,T)$  
has locally supported eigenfunctions. More precise: $\Oomega$ and  
$(\Omega^b,T)$ are {\it mutually locally derivable} (MLD). The  
equivalence concept of mutual local derivability for tilings was  
discussed in detail in \cite{BSJ}. This will all be discussed  
below. Our first result reads as follows.  
  
\begin{theorem}\label{mld}  
Let $\Oomega$ be a arbitrary DDSF. Then there exists a DDFS  
$(\Omega^b,T)$ and a random operator of finite range $(A^b_w)$  
such that $\Oomega$ and $(\Omega^b,T)$ are mutually locally  
derivable and $(A^b_w)$ has locally supported eigenfunctions with  
eigenvalue $E$ for every $\omega\in\Omega^b$. Moreover, $(A^b_w)$  
can be chosen to be the nearest neighbor laplacian of suitable  
graph.  
\end{theorem}  
  
\medskip

Note that for a selfadjoint random operator $A$ and bounded $Q\subset\RR^d$  
the restriction $A_\omega|_Q$ defined on $\ell^2(Q\cap\omega)$ has finite  
rank.  
Therefore, the spectral counting function  
$$  
n(A_\omega,Q)(E):=\#\{ \mbox{ eigenvalues of }A_\omega|_Q\mbox{  below  }E\}  
$$  
is finite and  
$\frac{1}{|Q|}n(A_\omega,Q)$ is the distribution function of the measure  
$\rho^{A_\omega}_Q$, defined by  
$$  
\langle \rho^{A_\omega}_Q,\varphi\rangle:=  
\frac{1}{|Q|}\mbox{tr}(\varphi(A_\omega|_Q))\mbox{  for  
}\varphi\in C_b(\RR). $$  
  
On a uniquely ergodic DDSF the measures $\rho^{A_\omega}_{Q_k}$  
converge in distribution to a measure $\rho^A$, called the  
integrated density of states (IDS) for any van Hove sequence $Q_k$  
as $k\to\infty$. This is described in \cite{LS2,LS3}. There one  
can also find an interpretation of the IDS as a certain trace on a  
von Neumann algebra. Now, we can state our main theorem.  
  
\begin{theorem}\label{main}  
Let $\Oomega$ be a strictly ergodic DDSF. Let $A$ be a selfajoint random operator  
of finite range.  
Then $E$ is a point of discontinuity of $\rho^A$  
if and only if there exists a locally supported eigenfunction of $A_\omega$ to $E$ for  
one (all) $\omega\in\Omega$.  
\end{theorem}  
  
\medskip  
  
\begin{remarks} {\rm  
\begin{itemize}  
\item[(1)] It rather straightforward to see that locally supported  
eigenfunctions lead to a discontinuity of the IDS. The more  
interesting part of the equivalence is discontinuities only happen  
in that way.  
\item[(2)] The theorem holds also in the tiling setting. Here a  
single tile of the original tiling will be replaced by 9 tiles of  
a new tiling which is MLD to the originally given one.  
\item[(3)] As pointed out already the theorem gives rise to a complete characterization of the phenomenon of  
locally supported eigenfunctions in quasicrystal settings (i.e. DDSF an tiling settings).  
\item[(4)] Let us emphasize that the integrated density of states is continuous in the  
case of almost periodic and random operators on lattices.  
Due to the more complex geometry this does not  
follow in the quasicrystal framework.  
\end{itemize}  
}  
\end{remarks}

\section{Preliminaries}  
In this section we will study the equivalence concept of MLD. Further  
we are going to construct a map, which maps any given DDSF  
to a DDSF which is MLD to the given one and where a random operator exists who has locally supported  
eigenfunctions.  
As well we state some  
tools which will be used later on to prove our results.  
  
\medskip  
As mentioned already the equivalence concept of MLD for patterns on tilings was discussed in \cite{BSJ}.  
We give the obvious definition for Delone dynamical systems of finite type.  
\begin{definition}  
Let $\Oomega$ and $(\Omega^b,T)$  a DDSF. A map  
$D:\Omega\to\Omega^b$ is called a local derivation map if there  
exists a radius $r_D>0$ such that  
$D(\omega)\cap\{x\}=(t+D(\omega))\cap\{x\}$ holds whenever  
$\omega\cap B(x,r_D)=(t+\omega)\cap B(x,r_D)$. In this case  
$(\Omega^b,T)$ is called locally derivable from $\Oomega$. Two  
DDSF $\Oomega$ and $(\Omega^b,T)$ are {\it mutually locally  
derivable} if $\Oomega$ is locally derivable from $(\Omega^b,T)$  
and vice versa (with a possibly different radius $r_D^\prime$).  
\end{definition}  
  
\medskip  
  
Note that the map $D$ is local in the sense that $D(\omega)\cap B(x,s)$ only depends on  
$\omega\cap B(x,s+2r_D)$.  
  
\begin{prop}  
Let $\Oomega$ be a DDSF, $D:\Omega\to\Omega^b$, $\omega\mapsto  
D(\omega)$ a local derivation map. Then $D$ is continuous with  
respect to the natural topology.  
\end{prop}  
{\it Proof.}  
This is immediate as $D$ is local and the topology is local in the sense of  
Lemma \ref{topology}.  
\qed  
  
\medskip  
  
What we want to do now is to put a well scaled local structure in  
a given Delone set whenever a certain pattern occurs. Let $\omega$  
be a Delone set and $P$ be a ball pattern class with $P\in\CalP$.  
Then, we define $\omega_P$ to be the set of all occurrences of $P$  
in $\RR^d$, i.e.  
\begin{equation}  
\omega_P\equiv\{t\in\RR^d: B(t,s(P))\wedge\omega =P\}.  
\end{equation}

Now, let $\Oomega$ be a DDSF and $r< r(\omega)$ for all  
$\omega\in\Omega$, $G$ be a finite graph with $V_G$ the set of  
vertices of $G$ contained in $\RR^d$. Furthermore let  
$\diam(G)=\tfrac{r}{21}$ and $V_G\subset B(0,\tfrac{r}{42})$. We  
use this finite graph to define a local derivation map by setting  
$ D_{P,V_G}(\omega)\equiv \omega\cup \{t+V_G:t\in\omega_P\}$ for  
$\omega \in \Omega$ and  
$\Omega^b:=\{D_{P,V_G}(\omega):\omega\in\Omega\}.$ Then,  
$$ D_{P,V_G}:\Omega\to\Omega^b,\quad  \omega \mapsto D_{P,V_G}(\omega)$$  
is a local derivation with inverse given  by the local derivation map  
$$H_{P,V_G}:\Omega^b\to\Omega,\quad H_{P,V_G}(\omega^b)=\{x\in\omega^b:\omega^b\cap B(x,\tfrac{r}{3})  
=\omega^b\cap B(x,\tfrac{r}{42})\}.$$  
Note that $\Oomega$ is also  
a local derivation of $(\Omega^b,T)$. Thus, $\Oomega$ and  
$(\Omega^b,T)$ are mutually locally derivable.

\begin{remarks} {\rm  
Let $\Omega$ and  $\Omega^b$ be as above. Then  
\begin{itemize}  
\item[(1)] If $\Oomega$ is a DDSF then $(\Omega^b,T)$ is a DDSF as well.  
\item[(2)] If $\Oomega$ is a uniquely ergodic DDSF, the same holds  
for $(\Omega^b,T)$.  
\item[(3)] The frequency of $G$ in $\Omega^b$ is the  
same as the frequency $\nu(P)$ of $P$ in $\Omega$.  
\end{itemize}  
}  
\end{remarks}  
  
\medskip  
  
The following two ingredients are essential for the proof of  
Theorem \ref{main}. The first one is one of the main results from  
\cite{LS3}. It relies on a strong ergodic type theorem proven  
there (see \cite{Len} for a study of uniform ergodic theorems in  the one dimensional case).

\begin{thm}\label{ids} Let $\Oomega$ be an aperiodic strictly ergodic DDSF.  
  Let $A$ be a selfadjoint operator of finite type and $(Q_k)$ be a van Hove sequence  
  in $\RR^d$. Then, the distribution functions of  
  $\rho_{Q_k}^{A_\omega}$ converge uniformly to the distribution function of the measure  
  $\rho^A$ and this convergence is uniform in $\omega\in \Omega$.  
\end{thm}  
  
The second one is a well known dimension argument from linear  
algebra which we will state for completeness reasons.  
  
\begin{prop}\label{dim}  
Let $H$ be a finite dimensional Hilbert space, $U,V$ subspaces of $H$ with  
$\mathrm{dim}\, U> \mathrm{dim}\, V$ then $\mathrm{dim}\,V^\perp\cap U >0$.  
\end{prop}  
  
\section{Proofs}  
In this section we prove Theorem \ref{mld} and Theorem \ref{main}.  
  
\medskip  
  
{\it Proof of Theorem \ref{mld}.}  
To prove the theorem we start with the construction of a DDSF  
where a random operator of finite range $(A_\omega)$ exists which  
has locally supported eigenfunctions.  
The starting point is a small graph $G_{fin}=(V_{fin},E_{fin})$ and  
an eigenfunction $u_{fin}$ of the associated nearest  
neighbor  
laplacian. For definiteness sake consider Figure 1. The values of $u_{fin}$ are  
indicated near the corresponding vertices. Here the eigenvalue is $E=0$.  
  
%
%
\begin{figure}[ht]  
\begin{center}  
\hspace*{-.5cm}  
\includegraphics[scale=.75]{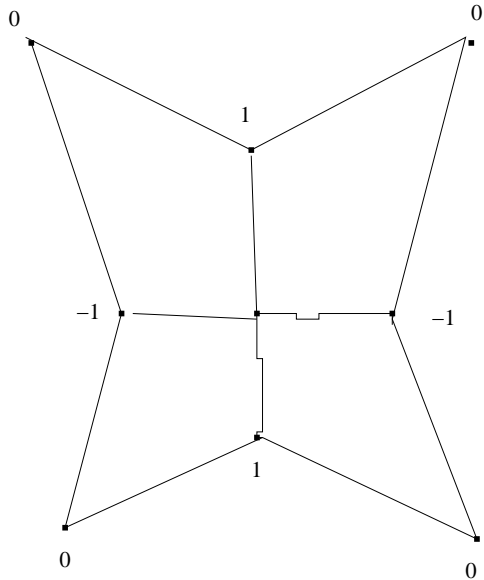}  
\caption{The finite graph $G_{fin}$.}  
\label{fig.1}  
\end{center}  
\end{figure}  
It is clear that whatever edges reach out of the four corners in a  
larger graph extending $G_{fin}$, the extension of $u_{fin}$ by  
$0$ to the larger vertex set will still constitute an  
eigenfunction of the laplacian on the large graph. It is now easy  
to implement this picture into a given DDSF. In fact, let  
$\Oomega$ be a DDSF and $P$ be a ball pattern. We use the local  
derivation map $D_{P,V_{fin}}$ discussed in the last section to put in $V_{fin}$ from above,  
scaled properly, whenever $P$ appears. It is obvious by the  
definition of $D_{P,V_{fin}}$, that this gives rise to a DDSF  
$(\Omega^b,T)$ which is locally derivable from $\Oomega$ and vice  
versa.  
  
Obviously we get a random operator $A^b$ with locally supported  
eigenfunctions by taking for  
$A^b_\omega$ the nearest neighbor laplacian on the copies of $E_{fin}$ in $\omega$  
and consistent  
matrix elements otherwise.  
  
\qed  
  
\medskip  
  
\begin{figure}[ht]  
\begin{center}  
\hspace*{-.5cm}  
\includegraphics[scale=.75]{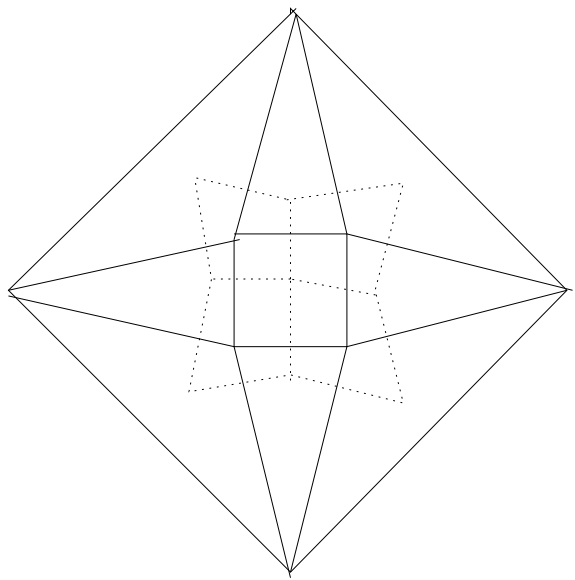}  
\caption{The tiling of $G_{fin}$.}  
\label{fig.2}  
\end{center}  
\end{figure}

\begin{remarks} {\rm  
\begin{itemize}  
\item[(1)] The simplest case of the construction made above is of course  
given by choosing $P=(\{x\},B(x,r))$ with $r<r(\omega)$. Then the  
graph $G_{fin}$ is glued at any point of the underlying Delone  
set. The corresponding $\ell^2$ space is just a direct sum (or  
tensor product) and that applies to the operators as well. Related  
constructions have been considered by \cite{SA} in the context of  
creation of spectral gaps.  
\item[(2)] For those who prefer tiling examples we now indicate how to view the  
construction above in this framework. Take a tiling dynamical  
system (see \cite{LS2,Sol2}) and replace one given tile $T$ by a  
suitable homeomorphic image of $T^b$ indicated in Figure  
\ref{fig.2}. We also indicated the next neighbor relations,  
showing that the resulting graph is just $G_{fin}$ above.  
\end{itemize}  
}  
\end{remarks}  
%
%
  
\medskip  
  
{\it Proof of Theorem \ref{main}.} We first show that the  
condition is sufficient. Let $u$ be an eigenfunction of  
$A_{\omega_0}$ associated to an eigenvalue $E$ with  
$\mathrm{supp}\, u\subset B(x,r)$ and $x\in\omega_0$. Then for any  
$\omega\in\Omega$ every copy of $P=B(x,r)\wedge\omega_0$ in  
$Q\wedge\omega$ adds a dimension to the eigenspace of $A_\omega|Q$  
belonging to the eigenvalue $E$. Let $\dot\sharp_P Q\wedge\omega$  
be the maximal number of disjoint copies of $P$ in  
$Q\wedge\omega$. Note that  
$\tfrac{|B(0,3r+r(\omega))|}{|B(0,r(\omega))|}=:C$ is a upper  
bound for the number of points (and therefore the maximal number  
of copies of $P$) in $B(0,3r)\cap\omega$.  
  
This gives by a direct combinatorial argument that  
$$\dot\sharp_P Q\wedge\omega\geq\frac{1}{C}\sharp_P Q\wedge\omega.$$  
  
Thus, for arbitrary $\epsilon>0$  
$$  
\frac{\tr(\chi_{(-\infty,E-\epsilon)}(A_\omega|_Q)}{|Q|}\le \frac{\tr  
(\chi_{(-\infty,E+\epsilon)}(A_\omega|_Q)}{|Q|}- \frac{1}{C}  
\frac{\sharp_P\omega\wedge Q}{|Q|} .  
$$  
Setting $Q=Q_k$ with $Q_k$ from a van Hove sequence and letting  
$k$ tend to infinity, we get that $\rho^A (E-\epsilon)\le \rho^A  
(E+\epsilon) - \tfrac{\nu(P)}{C}$. As $\epsilon>0$ is arbitrary  
and $\nu(P)>0$ the desired implication follows.  
\medskip  
  
Next we show the converse implication. Let $\tilde E$ be a point  
of discontinuity of the function $E \mapsto \rho_A((-\infty,E])$  
and $(Q_k)$ an arbitrary van Hove sequence. We consider the  
distribution function $\frac{1}{|Q|}n(A_\omega,Q)$ of the measure  
$\rho^{A_\omega}_{Q_k}$. Proposition \ref{ids} shows, that  
$\frac{1}{|Q_k|}n(A_\omega,Q_k)$ converges w.r.t. the supremum  
norm to the function $E \mapsto \rho^A((-\infty,E])$. Thus, for  
large $k$ the jump at $\tilde E$ of the function  
$\frac{1}{|Q_k|}n(A_\omega,Q_k)(E)$ does not become small. More  
precisely we get  
$$  
\mathrm{dim}\left(\mathrm{ker}\,(A_\omega|_{Q_k}-\tilde  
E)\right)=\lim_{\epsilon \to 0}(n(A_\omega,Q_k) (\tilde E  
+\epsilon)-n(A_\omega,Q_k)(\tilde E -\epsilon))\geq c|Q_k|  
$$  
for a $c>0$ and all $k\in\mathbb N$. Now let $\partial_{2r_A}  
Q_k\equiv \partial^{2r_A}Q_k\cap Q_k$ denote the inner boundary of  
range $2r_A$ of $Q_k$. For a van Hove sequence $(Q_k)$ we have  
\begin{eqnarray*}  
\mathrm{dim}\quad\ell^2(\partial_{2r_A} Q_k)  
&=& \sharp\{x\in\RR^d:x\in\partial_{2r_A} Q_k\cap\omega\} \\  
&\leq&  \frac{|\partial_{2r_A+r(\omega)} Q_k|}{|B(0,r(\omega))|}\\  
&=&\epsilon_k\cdot\frac{1}{|B (0,r(\omega))|}\cdot|Q_k|  
\end{eqnarray*}  
with a suitable $\epsilon_k$ which tends to $0$ for $k\to\infty$.  
For $k$ large enough we get that  
$c>\epsilon_k\cdot\tfrac{1}{|B(0,r(\omega))|}$. Thus, for large  
$k$ the inequality  
$$  
\mathrm{dim}\left(\mathrm{ker}\,(A_\omega|_{Q_k}-\tilde  
E)\right)>\mathrm{dim}\quad\ell^2(\partial_{2r_A} Q_k).  
$$  
holds. Now let $W_k$ the projection onto the inner boundary of range $2r_A$  
of $Q_k$. Then Propositon \ref{dim} shows that there exists an eigenfunction  
$f$ of $A_\omega$  
such that $W_kf=0$ for $k$ large enough.  
\qed  
  
\medskip  
  
\begin{remark}  
{\rm Let the conditions be as above. Then $E$ is an  
infinitely degenerate eigenvalue of $A_\omega$ for every $\omega\in\Omega$.  
The integrated density of states has a jump at $E$ whose height is at least  
$C^{-1} \nu(P)$.}  
\end{remark}  
  
\medskip  
  
{\bf Acknowledgements:}  
The authors would like to thank Uwe Grimm for helpful comments on the  
physics literature.

\end{document}